\documentclass[12pt]{iopart}

\usepackage{comment}
\usepackage{cite}
\usepackage{graphicx}% Include Fig. files
\usepackage{dcolumn}% Align table columns on decimal point
\usepackage{bm}% bold math
\usepackage{hyperref}% add hypertext capabilities
\usepackage[export]{adjustbox}
\usepackage[T1]{fontenc}
\usepackage{float}
\usepackage{todonotes}
\usepackage[normalem]{ulem}
\usepackage{color,xcolor}
\hypersetup{colorlinks=true,citecolor=blue,linkcolor=blue,filecolor=blue,urlcolor=blue,pdfstartview=FitH,pdfpagemode=UseNone}

\usepackage{array}
\newcolumntype{P}[1]{>{\centering\arraybackslash}p{#1}}

\begin{document}

\title{Edge disorder and magnetism in graphene nanoribbons: an inverse modelling approach}% Force line breaks with \\

\author{Shardul Mukim$^{1,2}$, Meric E. Kucukbas$^{1}$, Stephen R. Power$^{3}$ and Mauro S. Ferreira$^{1,2}$ }
\address{$^1$ School of Physics, Trinity College Dublin, Dublin 2, Ireland}
\address{$^2$ Advanced Materials and Bioengineering Research (AMBER) Centre, Trinity College Dublin, Dublin 2, Ireland}
\address{$^3$School of Physical Sciences, Dublin City University,  Glasnevin, Dublin 9, Ireland}
\ead{ferreirm@tcd.ie, stephen.r.power@dcu.ie}

\date{\today}

\begin{abstract}

%Fabrication of Graphene nanodevices is usually associated with 
It is difficult to completely eliminate disorder during the fabrication of graphene-based nanodevices. 
From a simulation perspective, it is straightforward to determine the electronic transport properties of disordered devices if complete information about the disorder and the Hamiltonian describing it is available. 
However, to do the reverse and determine information about the nature of the disorder purely from transport measurements is a far more difficult task.
In this work, we apply a recently developed inversion technique to identify important structural information about edge-disordered zigzag graphene nanoribbons. 
The inversion tool decodes the electronic transmission spectrum to obtain the overall level of edge vacancies in this type of device.  
We also consider the role of spin-polarised states at the ribbon edges and demonstrate that, in addition to edge roughness, the inversion procedure can also be used to detect the presence of magnetism in such nanoribbons. 
We finally show that if the transmission for both spin orientations is available, for example by using ferromagnetic contacts in a transport measurement, then additional structural information about the relative concentration of defects on each edge can be derived. 
\end{abstract}

\section{Introduction}
\label{sec:intro}

Graphene provides a unique zero band-gap electronic structure which has been the basis of numerous electronic devices\cite{neto2009electronic,park2021tunable,fet_bennett}.  
As a result, it has been the centre of attention of material scientists for more than two decades. 
Graphene nanoribbons (GNRs) are narrow strips of graphene which exhibit interesting electronic and physical properties \cite{son2006half,yano2019quest,cresti2008charge,wakabayashi1999electronic,cresti2007numerical,tombros2011quantized, baringhaus2014exceptional, caridad2018conductance}.
Chemical and structural alteration allows the tuning of GNRs to display insulating or metallic behavior \cite{son2006energy,han2007energy}. 
There are two fundamental edge geometries, namely armchair and zigzag edges.
In this article, we will focus on the latter, {\it i.e.} zigzag-edged graphene nanoribbons (ZGNRs).
ZGNRs can give rise to local magnetic moments at their edges, which combined with long spin-diffusion length in graphene, make these systems particularly promising in the realm of spintronics \cite{yazyev2010emergence, han2014graphene,liu2020spintronics,gregersen2017nanostructured,pedersen2008graphene,el2020progress,magda2014room}.
In fact, ZGNRs display spin-polarized transport channels near their edges, which can be further harnessed using electric fields and/or geometry effects to induce half-metallic behavior and spin-filtering functionalities\cite{aprojanz2018ballistic,sun2020coupled,jiang2020topology,ozaki2010dual}.

Huge experimental progress has been made in the synthesis and characterisation of high quality nanoribbons\cite{celis2016graphene, yano2019quest, chen2020graphene}.
Top-down lithographic etching\cite{han2007energy,tapaszto2008tailoring,datta2008crystallographic, caridad2018conductance}, bottom-up surface growth\cite{cai2010atomically, huang2012spatially, liu2015toward, basagni2015molecules, ruffieux2016surface}, unzipping of carbon nanotubes\cite{jiao2009narrow, jiao2010facile}, and epitaxial growth on silicon carbide sidewalls\cite{baringhaus2014exceptional, aprojanz2018ballistic} are a few examples of the existing techniques used to fabricate graphene nanoribbons.
While it is possible to grow or etch graphene structures with some degree of control over their edge geometry, the presence of disorder and edge roughness remains a very common artefact in most experimental nanoribbon systems\cite{kobayashi2006edge, niimi2006scanning, tapaszto2008tailoring,han2010electron, bennett2013bottom,pizzochero2021edge,pizzochero2021quantum}.

The transport and magnetic properties of GNRs depend heavily on the orientation and precision of the edges.
The presence of edge disorder can lead to Anderson localisation and transport gaps\cite{cresti2008charge, evaldsson2008edge, mucciolo2009conductance, ihnatsenka2009conductance}.
The presence of an edge vacancy can also disrupt the local magnetisation in its vicinity \cite{huang2008suppression,yazyev2011theory}. For a sufficient concentration of edge defects, it has been reported that the magnetic characteristics of ZGNRs can be completely quenched \cite{huang2008suppression, kunstmann2011stability}.
Scanning tunnelling microscopy and spectroscopy have revealed experimental proof of the edge states and associated magnetism by probing the local electronic structure \cite{ruffieux2016surface,tao2011spatially,magda2014room}. 
The observed states are usually compared to the theoretical simulations \cite{talirz2017surface,nguyen2017atomically, mishra2020topological}, so that verification of ZGNR magnetism is often indirect.
Characterising and understanding the role of different defect types and concentrations, and in particular how they effect both transport and local magnetism, is a vital step towards GNR-based electronics and spintronics.
Advancements in experimental techniques, computational methods, and theoretical models must work in parallel to  unlock the potential of ZGNRs for various applications.

Motivated by the desire to engineer the magnetic properties of GNRs, a number of studies have considered the  magnetic profiles generated by imperfect edges \cite{meric-prb, wimmer2008spin, yazyev2011theory, pizzochero2021edge}.
Such studies typically rely on solving the Schrodinger equation for Hamiltonians that reflect specific edge profiles. 
Assuming that the Hamiltonian associated with a GNR device is fully known, the task of finding the property of interest, {\it e.g.}, the spin-polarised conductance, is a straightforward one. 
But what if we approach this same task in reverse, {\it i.e.}, by trying to infer the edge quality starting from a given magnetic or transport property? 
In this article, we demonstrate an \emph{inverse problem} (IP) technique to characterise the level of edge disorder in ZGNRs using electronic transmission data. 
IP methods attempt to decode information about the causal factors of a system which are hidden in its observable signatures.
While they are at the heart of various imaging techniques\cite{medical, fwi, tromp2008spectral, sonar}, IP methods are not yet that common in the quantum realm. 
In one way or another, IP techniques involve comparing an input signal with the corresponding response.
In our case, these correspond to the parameters describing disordered ZGNRs and the simulated conductance spectra, respectively.

Individual defects have characteristic scattering effects that allow them to be identified by their transport signatures.
However, this becomes more challenging in the case of extended disorders, where a much wider range of complex multiple scattering events can occur. 
In this case, the transport response is sensitive not just to the type and concentration of defects, but to both the absolute and relative positioning of every defect in the system.
Decoding transport signals of this type, without any prior knowledge of the defect type, concentration or distribution, requires probing a vast phase space of disorder and GNR parameters.
The challenge lies in identifying the most suitable Hamiltonian, and the associated parameter values, which generate a match to the input signal - something that can be extremely laborious.
A naive approach to derive information about the concentration and spatial distribution of vacancies in a disordered system would be to compare the transmission signal of that system with those of several known disorder configurations in the hope of finding a match. 
However, the number of combinations required in practice to achieve this is far too high, proving this strategy impractical. 
The inversion technique introduced in Ref. \cite{shardul} accelerates this process by combining configurationally averaged quantities with the ergodic principle.
More specifically, the ergodic hypothesis assumes that a running average over a continuous parameter, in this case energy, upon which the transmission coefficient depends is equivalent to sampling different impurity configurations.
This allows the parameters of disordered structures to be identified from seemingly noisy conductance responses.

In this work, we extend the inversion methodology from Ref. \cite{shardul} to the case of edge-disordered ZGNRs to determine whether information about edge roughness can be reliably extracted from transport measurements of a specific device. 
In Section \ref{sec:method}, we introduce the spin-polarised and unpolarised Hamiltonians for ZGNR devices and the Landauer-Buttiker formalism used to calculate the transport response. 
Following that, in Section \ref{sec:results}, we introduce the inversion technique and demonstrate how it can effectively resolve both the degree of disorder in a ribbon and the presence of magnetic edges.  
Finally, in Section \ref{sec:asymm}, we demonstrate that the method can be extended to detect more subtle details of the disorder, such as the asymmetry between the two edges of the ribbon, when more complex spin-polarised transport measurements are included.

\section{Transport Properties}
\label{sec:method}

\begin{figure}
    \centering
    \includegraphics[width=0.8\linewidth]{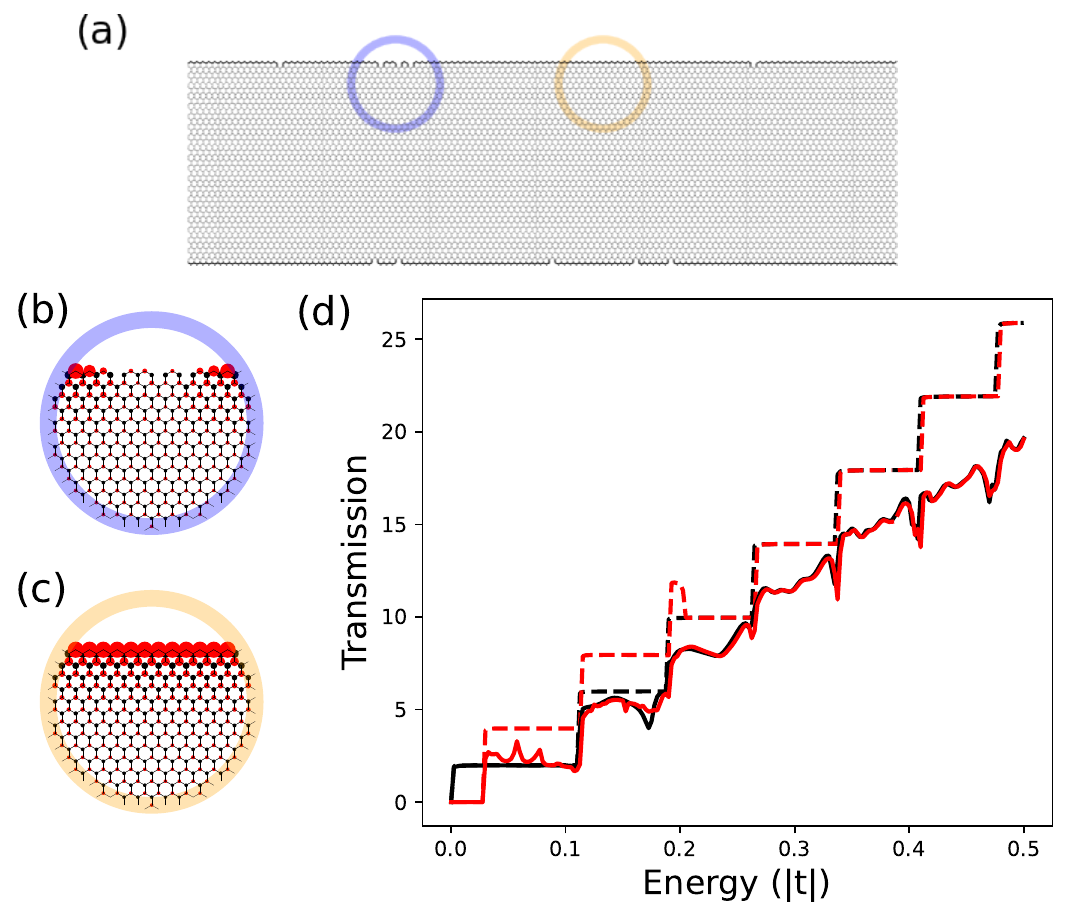}
    \caption{(a) Schematic of an edge-disordered 40-ZGNR with five vacancies on each edge ($N_T = N_B = 5, N=10$). The blue and orange circles highlight edge regions with and without vacancies, respectively. (b) and (c) show the magnetic moment profiles in these regions, with the colour and size of the symbols representing the sign and magnitude of the moment on each carbon site. (d) Transmission as a function of energy for the system in (a) shown by solid lines for both the  $U=0$ (black, non-magnetic) and $U=1.33|t|$ (red, magnetic) cases. The dashed lines show the corresponding transmissions for pristine ribbons (e.g. $N=0$). 
    }
    \label{fig:device}
\end{figure}

In this work, we consider electronic transport through pristine and edge-disordered nanoribbon segments such as those illustrated in Fig.~\ref{fig:device} (a) and (b).
Although the inversion methodology works for multi-terminal devices \cite{MUKIM2022360}, for simplicity, we consider a two-terminal setup with left and right leads connected to a GNR segment of width $W$ and length $L$.
The inversion process requires configurational averages over ensembles of systems with different impurity levels, and we begin by generating a large number of disordered ZGNR device geometries. 
In all cases the device leads are pristine ZGNRs, and only edges in the central part of the device contain disorder.
In particular, rough edges in the central device region result from vacancies introduced on both sides of the ribbon by randomly removing fixed numbers of atoms, $N_T$ and $N_B$, from the top and bottom edges respectively.
The total number of vacancies in a configuration is $N = N_T + N_B$.
We generated 50 different disorder realizations for each possible $(N_T, N_B$) pairing with $1 \le N_T, N_B \le 10$.
These were then split into training and test sets to generate configurational averages and validate the inversion procedure respectively.

We model the device using a single-orbital, nearest-neighbour tight-binding approach with a mean-field Hubbard term to include the effect of electron-electron interactions. 
We note that while spin-polarised density functional theory (DFT) would allow for a more complete description of the system, and perhaps a more accurate magnetic moment profile to be simulated, it is usually limited to small or periodic systems due to computational expense. 
The mean-field approach is found to be in good agreement with DFT calculations once suitable parameter values are chosen\cite{yazyev2010emergence}.
It is worth mentioning that the single-orbital simplification is by no means essential to the inversion approach, which can easily be generalised to more complex models. 
More specifically, the Hamiltonian is given by 
\begin{equation}
    {\hat H}^{\sigma} = 
    {\hat H}^{0} - \sigma\frac{U}{2}\sum_{i}m_{i}\,\hat{c}^{\dag}_{i\sigma}\,\hat{c}_{i\sigma}\,\,,
    \label{Hub-Ham}
\end{equation}
where $\hat{c}^{\dag}_{i\sigma}$  ($\hat{c}_{i\sigma}$) is the creation (annihilation) operator for electrons of spin $\sigma = \{\uparrow,\downarrow\}$ at site $i$, $m_i$ corresponds to the local magnetic moment at that site and $U$ represents the strength of the electronic interaction. 
%\soutM{$1.33\,t$, with $t = 2.7 {\rm eV}$ acting as the electronic hopping.} 
The operator ${\hat H}^0$ is the usual nearest-neighbour tight-binding Hamiltonian 
\begin{equation}
    {\hat H}^{0} =  \sum_{i,j,\sigma} \gamma_{ij}\,\hat{c}^{\dag}_{j\sigma}\,\hat{c}_{i\sigma}\,,
    \label{tb-Ham}
\end{equation}
with $\gamma_{ij}=t$ if $i$ and $j$ label nearest-neighbour sites but vanishes otherwise. The nearest-neighbour hopping $|t|  = - 2.7 {\rm eV}$ will be used as our energy unit.

Working with the Hamiltonian defined in Eqs.(\ref{Hub-Ham}) and (\ref{tb-Ham}) requires some care since the values of the local moments $m_i$ must be obtained through a self-consistent (SC) procedure which is by far the most onerous part of the calculation. 
In the presence of disorders, calculating the magnetic moments for an ensemble of long, disordered ribbons can be computationally very expensive. 
To speed up this process, some of the authors have previously developed a machine-learning approach~\cite{meric-prb} to bypass the SC calculation, allowing us to deal with many instances of large disordered systems. 
The machine learning model is trained over a vast phase space of disorder configurations of different edge roughness. 
This model has been extensively benchmarked against self-consistent calculations for a range of geometries, and both the moments themselves and quantities calculated from them (such as spin-polarised transmissions) are accurately reproduced~\cite{meric-prb}.
Once the local moments on each site are calculated, it is then straightforward to obtain the transmission $T_{L,R}$ across the structure
\begin{equation}
    T_{LR} = \mathrm{Tr} [\Gamma_{L}G^{r}_{LR}\Gamma_{R}G^{a}_{LR}]
    % \label{eq:lb}
    \label{conductance-exp}
\end{equation} 
Here, $G^{r}_{LR}(G^{a}_{LR})$ are retarded (advanced) Green's functions of the full system and $\Gamma_{L/R}$ are the line width matrices of the corresponding left (L) and right (R) leads connected to the system. 
The transmission in Eq.(\ref{conductance-exp}) corresponds to the measurable two-terminal conductance via the relation ${\cal G}_{LR} =  \frac{2e^2}{\hbar} T_{LR}$. 
To calculate the spin-polarised, and unpolarised transmission, we employ the recursive Green's function method to calculate the required matrix elements\cite{lewenkopf2013recursive, settnes2015patched, papior2017improvements}. 

To determine if an inverse model approach can be used to detect the presence of edge magnetic moments, it will also be necessary to compare transmissions with and without magnetic moments. 
This is done by selecting either the fully spin-dependent Hamiltonian ${\hat H}^{\sigma}$ from Eq.~(\ref{Hub-Ham}) or only the spin-independent hopping part ${\hat H}^{0}$ from Eq.~(\ref{tb-Ham}). 
This is also equivalent to using the full Hamiltonian with $U=1.33|t|$ for magnetic and $U=0$ for non-magnetic ribbons.
It will also be necessary to consider, either separately or combined, the transmissions of up- and down-spin electrons.
These will be denoted by $T_\uparrow$ and $T_\downarrow$ , with the total transmission for a magnetic ribbon denoted by $T_{\uparrow + \downarrow}$. 
The transmission for nonmagnetic ribbons $T_0$ has equal contributions from both spin orientations.

It is instructive to first consider the case of pristine ribbons, {\it i.e.}, in the absence of any vacancies. 
The dashed lines in Fig.~\ref{fig:device}(d) show the transmission spectra of a pristine 40-ZGNR ribbon using both the non-magnetic (NM, black curve) and magnetic (M, red curve) Hamiltonians. 
The main distinctions between the two cases are that magnetic ribbons open a small band gap $E_{G}$ at low energies and have an enhanced transmission over a small energy range beyond the gap.
Only negligible differences between the two cases exist at higher energies. 
Both key differences can be understood in terms of a spin-splitting of the flat band that occurs at $E=0$ in ZGNRs \cite{son2006energy}. 
Antiferromagnetic coupling between the moments at opposite edges determines the magnitude of $E_G$, with both decreasing for wider ribbons. 
Higher-order subbands are largely unchanged, as is the dispersive segment of the lowest energy mode which continues to carry currents of both spin orientations through the central region of the ribbon. 
The previously flat portions of the band now also gain dispersion and provide an additional transport channel for each spin in the energy range $\frac{E_{G}}{2} < |E| \leq 0.2|t|$. 
These new transport channels lie along the ribbon edges, with dispersive states for each spin located on opposite sides of the ribbon. 
This enhancement of the transmission at low energies is a key signature of the presence of spin-polarised edge currents in ZGNRs. 
Bearing in mind that inversion is all about identifying key fingerprints in a signal, Fig. \ref{fig:device}(d) tells us that when looking for magnetic signatures we should focus on this energy range. 
The solid curves in Fig. \ref{fig:device}(d) correspond to a 40-ZGNR with a disordered region of length $L=100$ similar to that shown schematically in Fig. \ref{fig:device}(a). 
In both cases, a suppression of transmission relative to the corresponding pristine case is observed, with significant differences between the NM and M cases again observed in the same low energy window.

\section{Inverse Modeling: Detecting Disorder and Magnetism}
\label{sec:results}

We now move on to investigate what geometric information can be retrieved from transport measurements of disordered samples. 
In this section we only consider systems with $|N_T - N_B| \le 1$, \emph{i.e.} with approximately symmetric distribution of vacancies on the top and bottom edges. 
We generate a training and test sets containing 44 and 6 disordered instances, respectively, for all possible impurity numbers $2 \le N \le 20$. 
First, we demonstrate how the inversion technique can be used to decode the total transmission spectrum of either a non-magnetic or magnetic ribbon. 
A system with a small number of vacancies ($N=5$) is chosen randomly from the test set to serve as our reference system. 
This can be thought of as a proxy for an experimental realisation, for instance, and the Hamiltonian describing it will be hereafter referred to as the {\it parent} Hamiltonian. 

We first consider the non-magnetic case, and take the energy-dependent transmission $T_0^p(E)$ associated with the reference structure, calculated within the Landauer-Buttiker formulism using the parent Hamiltonian, as the input signal to the inversion tool. 
Without using any other information about the reference system, we will test whether the inversion method can be used to determine $N$ -- the total number of vacancies present. 
Keeping in mind that the objective is to derive information about the structural composition of the GNRs, the inversion procedure introduced in \cite{shardul} provides an efficient way of combining configurationally averaged (CA) values of the transmission coefficients with the ergodic principle. 
In mathematical terms, this is done through the so-called \emph{misfit function} defined as \cite{shardul}
\begin{equation}
\chi_0(N)={1 \over {\cal E}_+ - {\cal E}_-}\,\int_{{\cal E}_-}^{{\cal E}_+} dE \,  \left(T_0^p(E) - \langle T_0(E, N) \rangle\right)^2\,\,,
\label{eq:misfit}
\end{equation}
where the integration limits $ {\cal E}_-$ and ${\cal E}_+$ are arbitrary energies which can be chosen to suit the specific problem. 
$\chi_0(N)$ can be interpreted as a functional that measures the deviation between the input transmission of the parent configuration $T^p_0(E)$ and its configurationally averaged counterpart  
\begin{equation}
\langle T_0(E,N) \rangle = {1 \over J} \sum_{j=1}^J T_0^j(E, N)\,,
\label{CA}
\end{equation}
where $T_0^j(E, N)$ is the transmission coefficient for a specific configuration $j$ with $N$ impurities in the training set, and $J=44$ is the total number of different disordered configurations for each impurity concentration in the training set. 
% We note that the random configurations in the ensemble are generated and have their transmissions calculated beforehand.
We emphasise that the reference system we are attempting to decode is from the test set, and is not used to generate the configurational average.
While both $T_0^p(E)$ and $\langle T_0(E,N) \rangle$ in the integrand above are functions of energy, the latter is also a function of the number of vacancies, $N$.

\begin{figure}
    \centering
    \includegraphics[width=\columnwidth]{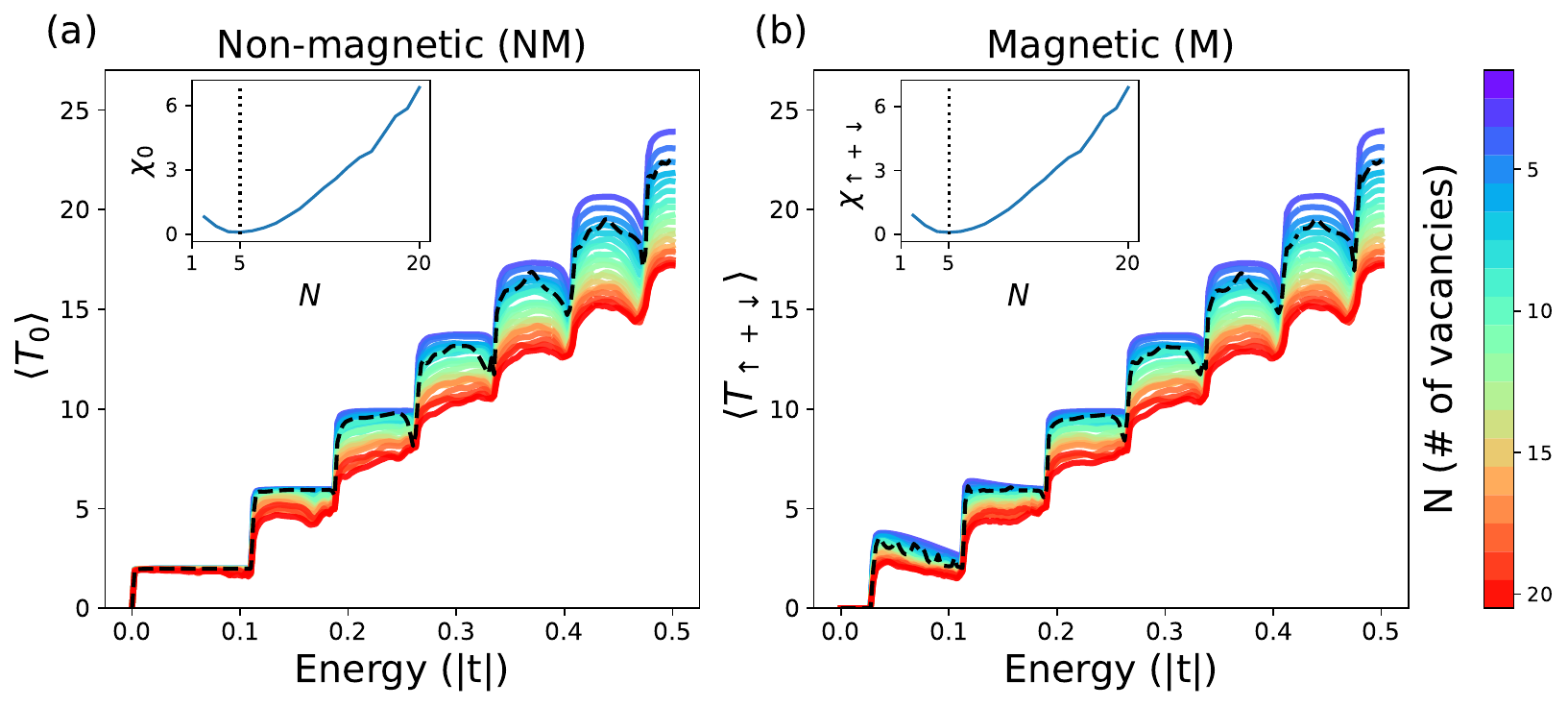}
    \caption{ Configurationally averaged transmission spectra for (a) non-magnetic and (b) magnetic ribbons with edge disorder. The colour-coded plots in each panel are configurational averages for different numbers ($N$) of edge vacancies. 
    The black dashed line in each panel is a test input (parent) signature $T_p(E)$ , corresponding to $N=5$ impurities. The insets in each panel show the corresponding misfit functions for this transmission, each of which show a minimum at $N=5$. 
    }
    \label{fig:transmission}
\end{figure}

The colour-coded curves in Fig.~\ref{fig:transmission}(a) depict the configurationally averaged transmissions $\langle T_0(E, N) \rangle$ for non-magnetic ribbons with different numbers of vacancies.
Each curve corresponds to an average over $J=44$ configurations. 
The curves form clear coloured bands, which indicate a reduction in transmission as N increases. 
As expected, the violet-colored bands, associated with a very low number of vacancies, resemble the staircase-like shape of the corresponding pristine curves in Fig.~\ref{fig:device}(d). 
The black dashed line indicates the transmission of the reference system $T^p_0(E)$ which we wish to invert.
By superimposing the dashed line with the colored bands, it is visually possible to estimate the number of vacancies contained in the parent system. 
However, a much more quantitatively accurate procedure is to graph the misfit function $\chi$. 
In fact, when plotted as a function of $N$, the misfit function displays a very distinctive minimum at a value that is likely to correspond to the real vacancy number (concentration) in the system. 
This can be seen in the inset of Fig.~\ref{fig:transmission}(a), where the curve indicates a minimum at the true value $N=5$. 
Fig.~\ref{fig:transmission}(b) repeats the procedure for magnetic ribbons. 
In this case the total transmission $T_{\uparrow + \downarrow}$ is used for both the reference system and to calculate the configurational averages required in the corresponding misfit function $\chi_{\uparrow + \downarrow}(N)$. 

\begin{figure}
    \centering
    \includegraphics[width=1.0\columnwidth]{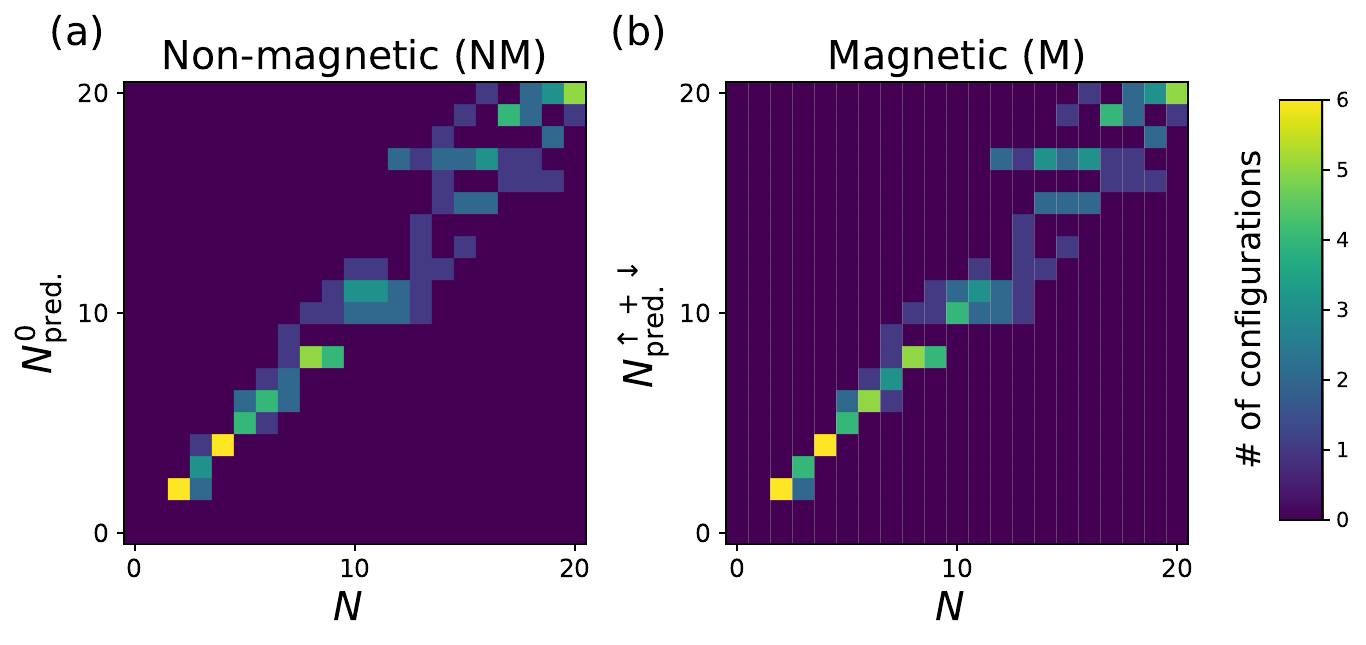}
    \caption{Comparison of the actual number of vacancies and the number of vacancies predicted by the inversion procedure for the entire test set. Electronic transmissions from (a) nonmagnetic and (b) magnetic ribbons yield similar levels of accuracy.}
    \label{fig:error}
\end{figure}

Despite differences in the magnetic and nonmagnetic transmissions at low energies, the misfit functions for both cases are very similar and both give an accurate prediction for $N$.
This suggests that, in either case, it may be possible to use $\chi(N)$ as an inversion tool to find the defects present on the edges of the ribbon from simple conductance measurements.  
It is important to note that no prior knowledge about the parent ribbon was necessary to identify the misfit function minimum. 
To test the method more thoroughly, we repeated the above procedure for all configurations in the test set.
The results are shown in the heatmaps in Fig.~\ref{fig:error}, which compare the actual number of vacancies present in the system to the prediction obtained from the inversion procedure for both (a) nonmagnetic and (b) magnetic ribbons. 
In both scenarios, the majority of configurations have their numbers of vacancies identified extremely accurately as the distributions in Fig.~\ref{fig:error} lie along the diagonal. 
We note that we expect discrepancies to increase as $N$ increases, as we note from Fig.~\ref{fig:transmission} that the configurational averages become more condensed, making it more difficult to differentiate between small differences in the total number of vacancies.

Energy integration plays a significant part in identifying the real impurity concentration because variations of $T(E)$ with energy in a single sample are equivalent to sample-to-sample fluctuations at a fixed energy, as seen in Refs. \cite{shardul, MUKIM2022360, duarte2024, duarte2024decoding}. 
Therefore, this energy integration is similar to vastly augmenting the number of disordered configurations taken into account. 
In fact, if we were to define the misfit function in Eq. (\ref{eq:misfit}) at only a single energy, it would not have any distinctive feature and would not lead to a successful inversion \cite{shardul}. 
However, if the integration limits span a small fraction of the relevant energy range, the misfit function acquires a very distinctive shape with minima located at the correct concentration, as seen in the insets of Fig.~\ref{fig:transmission}.
In this case, the bandwidth considered for the inversion ranges from ${\cal E}_-=0 $ and ${\cal E}_+ = 0.5t$, the full energy range shown in the transmission plots.

\begin{figure}
    \centering
    \includegraphics[width=1.0\columnwidth]{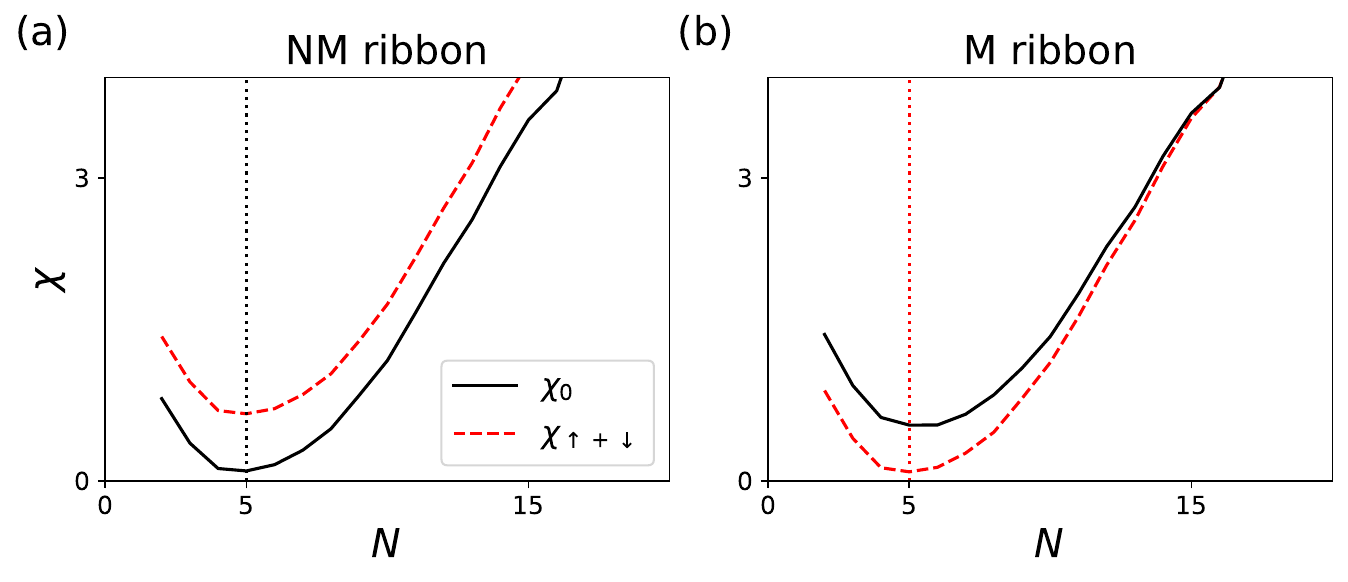}
    \caption{Comparison of nonmagnetic and magnetic misfit functions allows the presence of edge magnetism to be determined. In (a), the non-magnetic parent Hamiltonian is used, and the minimum of $\chi_0$ is below that of $\chi_{\uparrow + \downarrow}$, suggesting that no edge magnetism is present. Using a magnetic parent Hamiltonian in (b) yields the opposite result.
    }
    \label{fig:magtransmission}
\end{figure}

This leads to another critical question: is it possible to use the inversion tool to determine the presence of edge magnetism using only the transmission signature of an edge-disordered ribbon?
This could be easily done by simply calculating and comparing the misfit functions, $\chi_0$ and $\chi_{\uparrow + \downarrow}$, found using different sets of configurational averages -- those corresponding to magnetic and non-magnetic parent Hamiltonians. 
Fig.~\ref{fig:magtransmission} tests this idea by using both reference transmissions (dashed black lines) from Fig.~\ref{fig:transmission}(a) and Fig.~\ref{fig:transmission}(b) individually as input to both $\chi_0$ and $\chi_{\uparrow + \downarrow}$ mistfit functions. 
For a non-magnetic ribbon, both $\chi_0$ and $\chi_{\uparrow + \downarrow}$ have minima at $N=5$ in Fig.~\ref{fig:magtransmission}(a), but the minimum of $\chi_0$ is smaller, suggesting (correctly) that the system is non-magnetic. 
Fig.~\ref{fig:magtransmission}(b) shows that the minimum of $\chi_{\uparrow + \downarrow}$ is smaller if a magnetic reference transmission is used instead, indicating the presence of edge magnetism.
When this procedure is repeated across the test set, the presence of edge magnetism could be predicted with 100\% accuracy.
The simplicity of the technique can be used not only to derive fundamental information about the defects present in the system, but also it gives important information about the underlying Hamiltonian of the GNR.

\section{Inverse Modeling: Asymmetric Edges}
\label{sec:asymm}

Up to this point, we have inverted the total transmission of ribbons with or without magnetic edges in order to identify both the concentration of edge vacancies and the presence of edge magnetism. 
We will now demonstrate that the unique electronic properties of magnetic ribbons allow us to also decode information about the relative impurity distribution on both edges, \emph{i.e.} to determine not only the total number of vacancies $N$ but also the number of vacancies on the top ($N_T$) and bottom ($N_B$) edges individually.
Extra transmission channels are present at low energies in magnetic ZGNRs, as can be clearly seen by comparing the red and black-dashed curves in Fig.~\ref{fig:device}(d).
These additional channels emerge due to spin-splitting of the zero-energy flat band seen in non-magnetic ZGNRs, and are associated with dispersive states on either side of the ribbon which are strongly spin- and sublattice-polarised. 
The presence of edge-localised transport channels, with different spin orientations at each edge, suggest that spin-polarised transmission signatures should encode structural information about each edge that is lost when only the total transmission is considered.  

\begin{figure}
    \centering
    \includegraphics[width=\columnwidth]{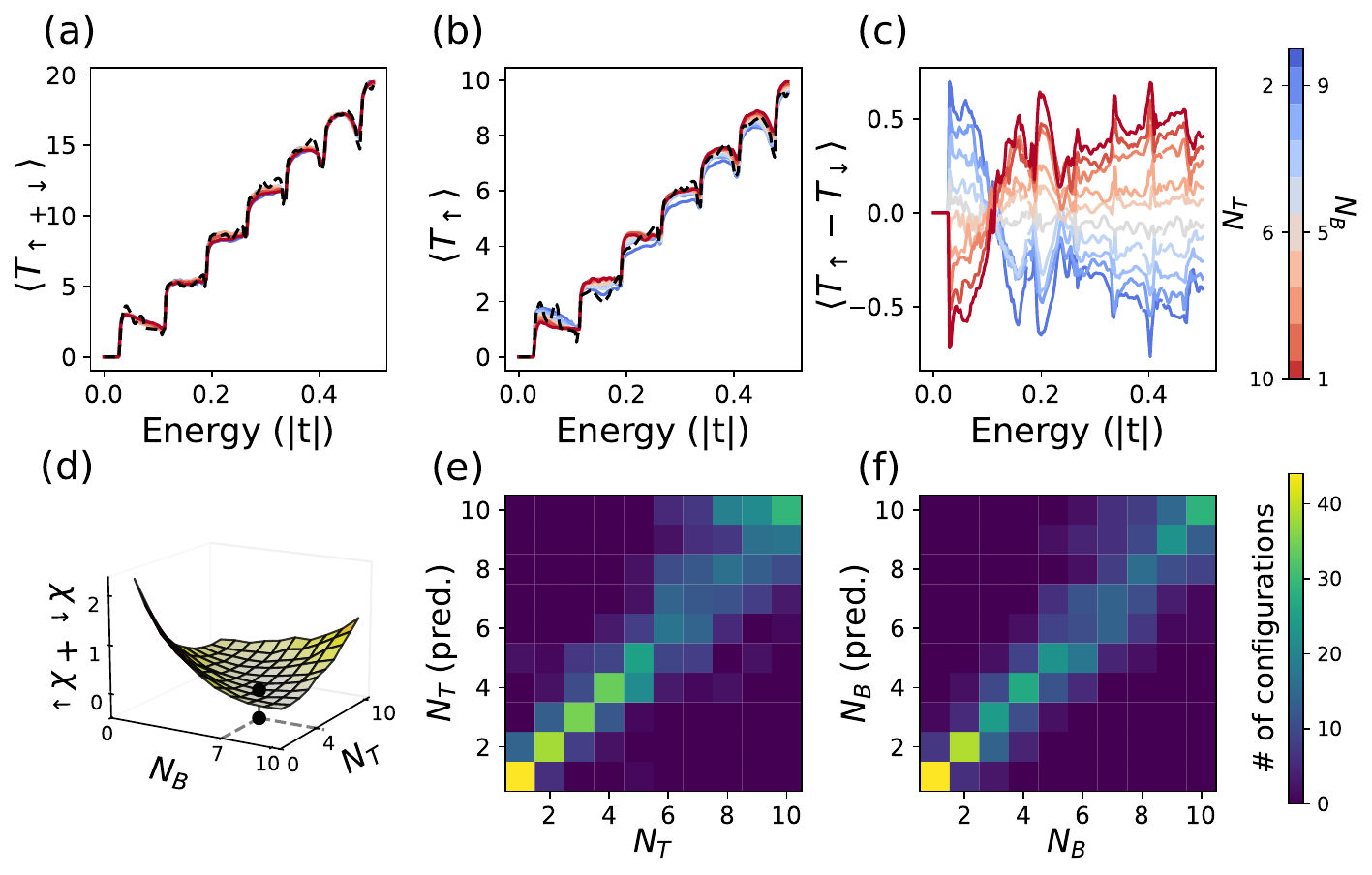}
    \caption{ (a) Configurationally averaged total transmissions $\langle T_{\uparrow+\downarrow}\rangle$ through disordered ZGNRs with the same total number of impurities ($N=11$). The different coloured curves, indicating different distributions of vacancies between the top and bottom edges, all coincide. 
    (b) More difference is seen between different vacancy distributions if only the averaged spin-up transmissions $\langle T_{\uparrow}\rangle$ are considered. (c). The effect of vacancy distribution is particularly clear in the pure spin current $\langle T_{\uparrow}-T_{\downarrow}\rangle$. 
    (d) A two-dimensional misfit function allows the number of vacancies on each edge to be decoded from the reference transmissions shown by dashed lines in panels (a) and (b). (e), (f) show the accuracy of these predictions for the entire test set. }
    \label{fig:spinuptransmission}
\end{figure}

For this investigation, we consider all $(N_T, N_B)$ pairings in our data set, and again generate training and test sets with 44 and 6 disordered instances for each $1\le N_T, N_B \le 10$.
To show the role that edge asymmetry has on transport properties, we first consider all configurations with $N=11$ vacancies in Fig.~\ref{fig:spinuptransmission}(a)-(c), with the different coloured curves representing training set configurational averages over different distributions of these impurities between the two edges, as indicated by the colour bar. 
The dashed black curve in Fig.~\ref{fig:spinuptransmission}(a) and (b) is a reference potential from the test set with $N_T=4$ and $N_B=6$.
Fig.~\ref{fig:spinuptransmission}(a) shows the total transmissions through these systems.
The curves for all vacancy distributions collapse onto each other, making it impossible to differentiate between them.
However, the curves begin to separate if we only consider the up-spin transmission, as in Fig.~\ref{fig:spinuptransmission}(b).
Here we note that, for example, within the first plateau the red curves have lower values than the blue ones, meaning that $T_\uparrow$ is more suppressed by vacancies on the top edge than the bottom edge in this energy window. 
This is because the new transmission channel for up-spin electrons in magnetic ribbons is located near the top edge of the ribbon and is more strongly affected by vacancies on this edge.
A similar, but reversed, behaviour occurs for $T_\downarrow$, as the associated transport channel is near the bottom edge of the ribbon. 
The effect of asymmetry is therefore extremely clear in the pure spin current $T_\uparrow$ - $T_\downarrow$, shown in Fig.~\ref{fig:spinuptransmission}(c). 
Indeed, the presence of a non-zero spin current requires an asymmetry between the up- and down-spin channels, which only occurs in our systems due to an interplay between the different spatial distributions of up- and down- spin wavefunctions and the different scattering profiles near each edge.

To decode the vacancy distributions, we define two-dimensional misfit functions, which depend separately on $N_T$ and $N_B$, for each spin channel 
\begin{equation}
    \chi_{\sigma}(N_T, N_B)={1 \over {\cal E}_+ - {\cal E}_-}\,\int_{{\cal E}_-}^{{\cal E}_+} dE \,  (T_{\sigma}(E) - \langle T_{\sigma}(E, N_T, N_B) \rangle)^2\,\,
    \label{eq:spinmisfit}
\end{equation}
where $\sigma = \uparrow$ or $\downarrow$.
We then use the combination $\chi_\uparrow + \chi_\downarrow$ as the final misfit function to determine the impurity distribution. 
We note that this quantity is the sum of misfits from two different spin-dependent measurements, and is therefore different from $\chi_{\uparrow + \downarrow}$,  used in Sec.~\ref{sec:results}, which is a misfit calculated from the total transmission and which averages out differences between the up- and down-spin transmissions.
Fig.~\ref{fig:spinuptransmission}(d) plots this misfit function for the reference transmission shown by dashed lines in Fig.~\ref{fig:spinuptransmission}(a) and (b), which has a minimum exactly at the point corresponding to the correct number of vacancies on each edge.  
Fig.~\ref{fig:spinuptransmission}(e) and (f) shows the results of applying this procedure to all 600 configurations in the test set, with very accurate predictions found for the number of vacancies on both edges. 
More details about how the accuracy achieved depends on the size of the dataset, and of the width and length of the disordered ribbons, can be found in \ref{sec:app}.

\section{Conclusion}
Our work demonstrates that electronic transport can be used to quickly characterize the level and distribution of edge vacancies in ZGNRs, as well as to determine whether or not magnetic moments have arisen along the zigzag edges.
This provides an additional tool, complementary to scanning tunneling microscopy and spectroscopy to probe GNR devices.
The inversion technique presented here can be used to procure important structural information about the system or about the underlying Hamiltonian that best describes it. 
The misfit function is capable of detecting not only edge roughness, but also edge magnetism, without availing of any additional characterization information aside from a transmission spectrum. 
With additional pieces of information, in the form of individual spin-up or spin-down transmission signatures which can be acquired using ferromagnetic contacts, the inversion procedure can also be used to identify the relative distribution of vacancies on both edges.

Finally, we note that while the method is developed around a relatively simple Hamiltonian and form of edge disorder, \emph{i.e.} a certain number of vacancies on each edge, it can easily be extended to more complex systems. 
The extent of stronger edge disorders may be more accurately characterised by a roughness parameter or the particular details of the etching procedure~\cite{mucciolo2009conductance, caridad2018conductance}, which a similar inversion procedure would be able to decode from fingerprints in transmission spectra. 
Alternatively, using \emph{ab initio} methods to obtain more detailed representations of individual and coupled defects would allow the effects of chemical functionalisation or local deformations to be included~\cite{yabusaki2019effects}. 
Finally, while we have focused on the antiferromagnetic alignment of moments on different sublattices and edges, a ferromagnetic alignment can be generated by varying the width or doping levels of the system\cite{magda2014room, chen2017width}. 
In this case, the inversion procedure would still be able to determine information about the total number of vacancies, but details about individual edges would be more difficult to decode.

\section*{Acknowledgments}
   The work of S.M. and M.S.F. emanated from research supported partly by a research grant from Taighde Eireann (former SFI) under grant no. SF12RC2278$\_$P2. 
      S.R.P. wishes to acknowledge funding from the Irish Research Council under the Laureate awards programme, and the gazelle computational facility in the School of Physical Sciences at DCU, which is supported by Intel Ireland.

\appendix
\section{Dataset and system sizes}
\label{sec:app}
In analogy with machine-learning approaches, the size of the training set should affect the quality of the predictions returned by the inversion procedure. In Fig.~\ref{fig:appendix}(a) we show how the prediction error depends on the number of configurations used to calculate the configurational average for each $(N_T, N_B)$ pair. 
The Root Mean Square Error (RMSE) on every $N_T$ and $N_B$ prediction made on the test set, i.e. all the predictions shown in Fig.~\ref{fig:spinuptransmission}(e) and (f),  is used as the error metric. 
A higher number of configurations will give smoother configurational averages for use in the misfit function in Eq.~(\ref{eq:spinmisfit}). 
The RMSE saturates quite quickly as the number of configurations used is increased, and no considerable improvement is expected beyond the $J=44$ configurations used throughout this work.
We note that the configurational average does not necessarily need to be completely converged for the method to work, as long as the variations from the fully-converged configurational average are smaller than those between different numbers of vacancies.

The electronic properties of nanoribbons are sensitive to their width~\cite{son2006energy} and the transport properties of disordered systems are sensitive to the length of the disordered region~\cite{mucciolo2009conductance}, so we now briefly consider how these factors affect the performance of the inversion procedure.
Fig. ~\ref{fig:appendix}(b) shows the RMSE calculated for three different ribbons widths, with the central 40-ZGNR case being that discussed in detail in the main paper. 
Fig. ~\ref{fig:appendix}(c) shows a similar analysis for three different ribbon lengths for the narrowest 20-ZGNR ribbon.
While the error in all  cases shown are of similar magnitude, we note that the predictions become slightly less accurate for narrower or shorter systems, where the vacancies are more concentrated. This suggests that for higher concentrations, the number of vacancies alone is not necessarily sufficient to characterise the edge disorder.

\renewcommand{\thefigure}{A\arabic{figure}}

\setcounter{figure}{0}

\begin{figure}
    \centering
    \includegraphics[width=\columnwidth]{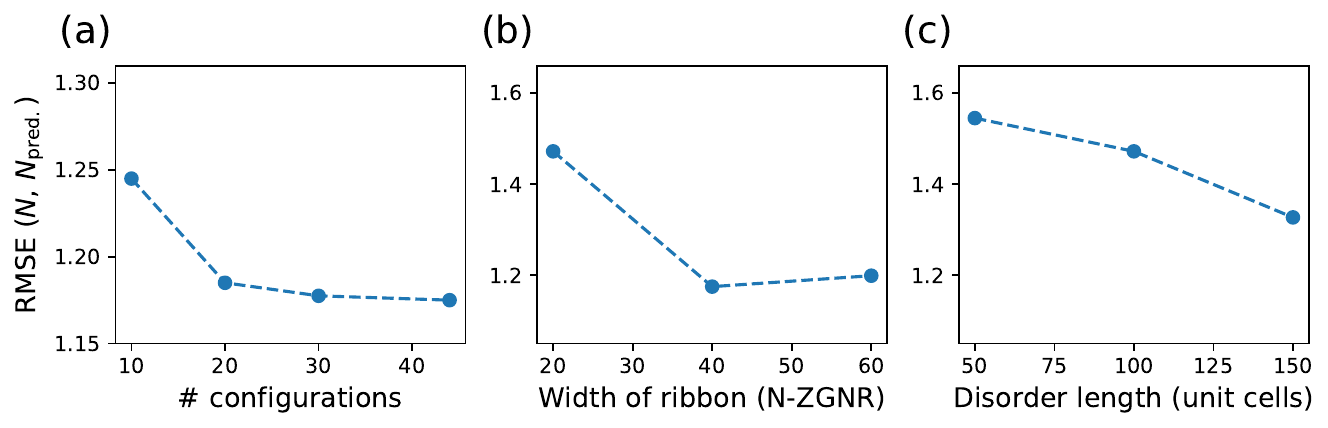}
    \caption{ Root mean square errors (RMSE) on the predicted number of vacancies on each edge, as determined using the approach outlined in Sec. \ref{sec:asymm}. The panels show the dependence of the RMSE on (a) the number of disordered configurations considered for each configurational average, (b) the width of the ribbon (for fixed $L=100$), and (c) the length of the disordered region (in unit cells, for a fixed width 20-ZGNR).   }
    \label{fig:appendix}
\end{figure}

--------------------------------
\section*{References}

% \bibliography{mybib}

% \

\end{document}